\documentclass[aps,pre,twocolumn,showpacs]{revtex4}
\usepackage{amsmath,amsfonts,bm, color,amssymb}
\usepackage{graphicx,epsfig,overpic,hyperref,ulem}

\newcommand{\refeq}[1]{Eq.(\ref{#1})}

\newcommand{\Rr}{\mbox{\it R}}
\newcommand{\Sr}{\mbox{\it S}}
\newcommand{\mw}{{\mathrm{mw}}}

\newcommand{\sigm}{{\sigma}}
\newcommand{\eps}{{\varepsilon}}
\newcommand{\out}{{\mathrm{m}}}
\newcommand{\ins}{{\mathrm{p}}}

\newcommand{\bE}{{\bm{E}}}

\newcommand{\bOmega}{{\bm{\Omega}}}

\begin{document}
\title{Quincke rotor dynamics in confinement: rolling and hovering}
\author{Gerardo E.~Pradillo\textit{$^{a}$},Hamid Karani\textit{$^{b}$}, Petia M.~Vlahovska\textit{$^{a,b}$}}
\affiliation{%
$^{a}$~Mechanical Engineering, Northwestern University, Evanston, IL 60208, USA\\
$^{b}$~Engineering Sciences and Applied Mathematics, Northwestern University, Evanston, IL 60208, USA. E-mail: petia.vlahovska@northwestern.edu }

\date{\today}

\begin{abstract}
The Quincke effect is  an electrohydrodynamic instability which gives rise to a  torque on a dielectric particle in a uniform DC electric field. Previous studies reported that a sphere initially resting  on the electrode rolls with steady velocity. We experimentally find that in strong fields the rolling becomes unsteady, with time-periodic velocity. Furthermore, we find  another regime, where the rotating  sphere levitates in the space between the electrodes.    
 Our experimental results show that the onset of Quincke rotation strongly depends on particle confinement and the threshold for rolling is higher compared to rotation in the hovering state. 
 \end{abstract}

\maketitle

\section{Introduction}

The spontaneous spinning of a dielectric sphere in a uniform DC electric field  was described over a century ago in the work of G. Quincke  \cite{Quincke:1896}. The phenomenon was studied sporadically  \cite{Melcher-Taylor:1969,Jones:1984, Turcu:1987, Lemaire:2002,  Lemaire:2005, Vlahovska:2019}, however in recent years it  is enjoying increasing  interest. 
An isolated sphere was found to undergo Lorenz chaotic rotations \cite{Lemaire:2005} and pairs of Quincke rotating spheres display intricate trajectories \cite{Das-Saintillan:2013, Dolinsky-Elperin:2012, Dommersnes:2016prop}.  A suspension of Quincke rotors can exhibit lower effective viscosity \cite{Cebers:2004, Lemaire:2008, Huang-Zahn-Lemaire:2011}  or increased conductivity \cite{Lemaire:2009b} compared to the suspending fluid. More complex electrorotation dynamic arises from field nonuniformity  \cite{Yi:2018} or nonspherical particle shape 
  \cite{Cebers:2000, Cebers:2002,Dolinsky-Elperin:2009, Brosseau:2017a}, for example,  
  {shape anisotropy created by chirality \cite{Das-Lauga:2019} or  deformation as in the case of an elastic   filament attached to a  sphere   \cite{Zhu-Stone:2019} converts the Quincke rotation into particle translation. }
  Drops while rotating can also deform and appear as if ``breathing"  \cite{Sato:2006, Salipante-Vlahovska:2010, Salipante-Vlahovska:2013, Ouriemi:2015, Vlahovska:2016review, Vlahovska:2019}.   Quincke rotation in complex media is affected by the medium structure. For example, in liquid crystals Quincke rotors orbit along circularly shaped smectic defects \cite{Jakli:2008, Lavrentovich:2016}. Quincke rotors initially resting on a surface roll with steady velocity.  Large populations of these so called Quincke rollers can self-organize and undergo 
 directed motion
  \cite{Bartolo:2013, Bartolo:2015, Belovs:2014, Lu:2018, Geyer:2018}, although heterogeneous medium may suppress the collective motion and destroy the  Quincke roller flocks \cite{Morin:2017a, Morin:2017b}. 
 
\begin{figure}[h]
\centerline{\includegraphics[width=\linewidth]{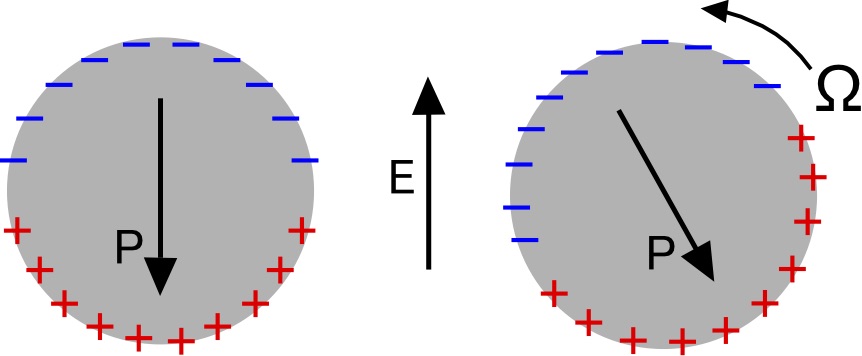}}
\caption{\footnotesize (left) Induced free charge distribution  for a sphere with $\Rr/\Sr<1$.  (right) Above a critical field strength $E>E_Q$ steady rotation 
in the plane perpendicular to the electric field ($\bOmega\cdot\bE=0$) is induced by the misaligned induced dipole  of the particle. 
}
\label{fig1}
\end{figure}

The Quincke effect arises from particle electric polarization, see Figure \ref{fig1} for an illustration of the mechanism.
Upon application of  an electric  field, mobile ions brought by conduction 
accumulate at the particle interface due to the difference  of  electrical conductivity, $\sigm$,  and permittivity, $\eps$ between the particle, $``\ins "$, and suspending, $``\out "$,  media.
In a uniform field, if the resulting induced dipole is antiparallel to the the applied field, {a spontaneous symmetry breaking occurs in strong fields.}
The theoretical analysis of this instability for a sphere in an unbounded domain predicts that the dipole adopts a steady tilt angle relative to the applied field direction above a threshold electric field    \cite{Jones:1984, Turcu:1987, Lemaire:2002}
 \begin{equation}
\label{quinckeE}
\begin{split}
E_Q^2=\frac{2\sigm_\out \mu_\out \left(\Rr+2\right)^2}{3\eps_\out ^2 (\Sr-\Rr)}\,, 
 \end{split}
\end{equation}
where
\begin{equation}
\Rr=\frac{\sigm_\ins}{\sigm_\out}\,,\quad \Sr=\frac{\eps_\ins}{\eps_\out}\,.
\label{ParameterRatios}
\end{equation} 
The resulting electric torque drives  rotation with rate $ \Omega_Q$, which increases with field strength
 \begin{equation}
\label{quinckeW}
\begin{split}
 \Omega_Q=\pm \frac{1}{t_{\mw}}\sqrt{\frac{E^2}{E_Q^2}-1}\,,\quad t_\mw=\frac{\eps_\out}{\sigm_\out}\left(\frac{\Sr+2}{\Rr+2}\right)
 \end{split}
\end{equation}
 \refeq{quinckeE} shows that rotation is possible only if the material properties are such that $\Rr/\Sr<1$.
 
Here we experimentally investigate the effect of confinement on the Quincke rotor dynamics. Our study is motivated by the fact that in the experiments with Quincke rollers \cite{Bartolo:2013, Bartolo:2015} the particles are sandwiched between electrodes and are rolling on the bottom surface.  
Since the Quincke effect is very sensitive to the suspending fluid conductivity we control it by adding surfactant. We find that the additive strongly influences the Quincke dynamics  and in addition to rolling,  
 we find a new regime of hovering, where the sphere lifts off the bottom surface and spins in the space between the electrodes.

\section{Methods and materials}

\subsection{Experimental setup}

\begin{figure}[h]
\centerline{\includegraphics[width=\linewidth]{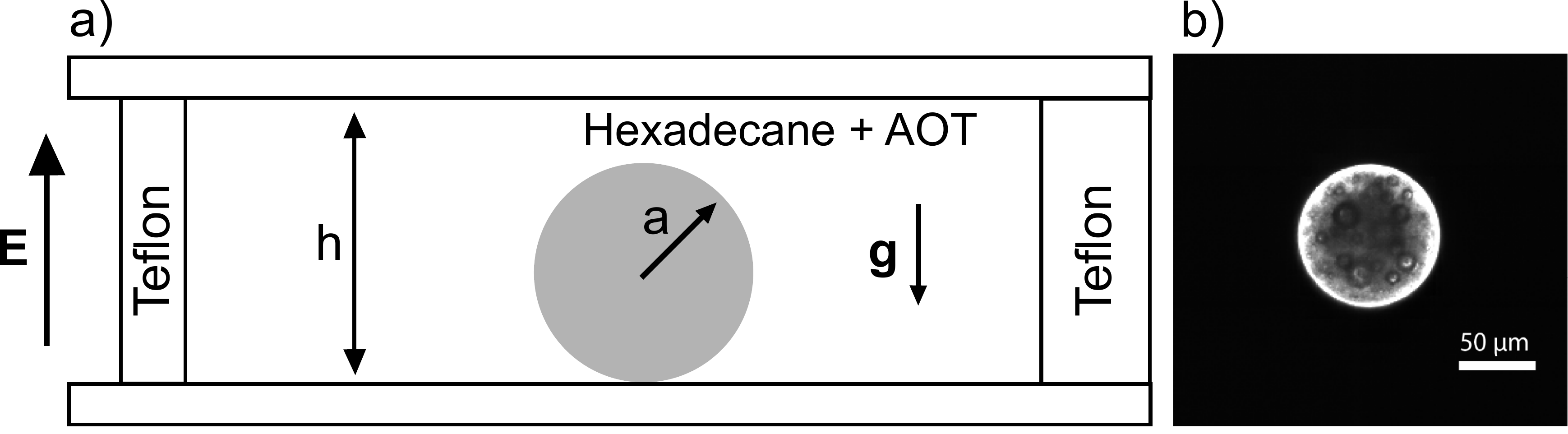}}
\caption{\label{fig:setup2} \footnotesize a) Side view of experimental setup. 
b) Image of single PMMA particle. Rotation rate is measured by tracing the defects.}
\end{figure}

The experimental setup consists of two indium-tin-oxide (ITO) coated  glass slides (Delta Technologies) separated by a  Teflon tape of thickness $h$, as shown in Figure \ref{fig:setup2}. The chamber area is 2~$\times$~2~cm$^2$. 
A single colloidal size poly-methyl methacrylate (PMMA) (Phosphorex) particle is placed on the bottom electrode of the chamber,  and then the chamber is filled with  hexadecane containing small amount of AOT(Dioctyl sulfosuccinate sodium salt) (Sigma Aldrich). 
 To create different rotor confinement, we study particles with  diameters  $d = 40 \mu m$ and $d = 100 \mu m$, and  a chamber with teflon spacer  $h = 120 \mu m\ \text{and}\ h = 240 \mu m$. We characterize the confinement by the ratio $d/h$, although the actual gap between between the electrodes is about 10-15 $\mu m$ larger than the tape height $h$ due to  fluid penetration between the tape and the glass.   A potential difference between the ITO-electrodes is applied using a high voltage amplifier (Matsusada). Observations are done using a Zeiss microscope. High speed camera (Photron) is used to record images of the particle, which are analyzed using a custom Matlab code, to extract the trajectory and velocity of the colloid.

The material properties for the particles and suspending fluid are listed in Table \ref{matprop}. 
 The electrical conductivity of the fluids is measured using a high-precision multimeter (BK Precision), following a similar procedure as in Sainis et al.\cite{Sainis:2008b}. The electric conductivity of the pure hexadecane was below the sensitivity of the multimeter, $10^{-12}$ S/m. 

\begin{table}[h]
\begin{center}
\resizebox{\columnwidth}{!}{%
\begin{tabular}{lcccccc}
\hline\hline
 &  Density& Dielectric  & Conductivity & Viscosity  \\
 & $\rho$ & constant & $\sigma$ & $\mu$ \\
Material &  (g/cm$^3$) & $\epsilon$ (-) & (S/m) & (Pa.s)\\
\hline
Hexadecane & 0.77& 2 &NA &$4.3\times 10^{-3}$ \\
PMMA & 1.18 &3.6 & $10^{-17}$ &NA \\
\hline \hline
\end{tabular}
}
\caption{Material properties of the particle \cite{Ouriemi:2015} and suspending fluid. }
\label{matprop}
\end{center}%
\end{table}

\subsection{Control of fluid conductivity}

The conductivity of the suspending fluid is controlled by adding AOT (Sigma Aldrich) to hexadecane (Sigma Aldrich). We use $0.05$, $0.10$, and $0.15$ M AOT solutions similar to  the compositions that have been used in previous studies of Quincke rollers  \cite{Bartolo:2013, Bartolo:2015, Belovs:2014, Lu:2018, Geyer:2018}. These concentrations are well above the Critical Micelle Concentration (CMC) for AOT, which is about $10^{-6}$ M \cite{Sainis:2008b}. The Quincke effect 
is not observed in pure (AOT-free) hexadecane and hexadecane with low (sub- and near-CMC) AOT concentrations {likely due to very low number of free charges, which renders the media effectively perfect dielectrics}. {AOT is also known to  charge colloidal particles  dispersed in oil \cite{Hsu:2005, Sainis:2008a, Eastoe:2013, Eastoe:2015}. In our system, the particle stays on the bottom electrode even upon reversal of the field polarity, which suggests that charging, if any, is negligible. }

AOT is hygroscopic and the original package, as delivered from the manufacturer, contains uncontrolled amount of moisture \cite{Sainis:2008a}. We remove this  water by drying the salt in a convection oven (Fisher Scientific) at 90 C for 24 hours. Controlled amounts of moisture to the AOT are introduced by placing the salt in a custom built humidity chamber for different periods of time.  We mix the salt with the hexadecane and wait for approximately 2 hours before carrying out experiments.
 The hexadecane-AOT solutions used in the experiments and their conductivities are listed in Table \ref{cond}. 
Notably, the solution conductivity has a non-trivial dependence on AOT and water content and different combinations of AOT and moisture can result in a similar conductivity.  
\begin{table*}[t]
\begin{center}
\begin{tabular}{ccc|ccc|ccc}
\hline\hline
\multicolumn{3}{c}{0.05 M AOT } & \multicolumn{3}{c}{0.1 M  AOT } & \multicolumn{3}{c}{0.15 M AOT } \\
Water ($\%_{wt}$) && $\sigma$ (S/m) & Water ($\%_{wt}$) && $\sigma$ (S/m) & Water ($\%_{wt}$) && $\sigma$ (S/m) \\
\hline 
0 && 4.91 $\times 10^{-9}$  & 0 && 1.26  $\times 10^{-8}$ & 0 && 2.32 $\times 10^{-8}$\\
1.86 && 6.87  $\times 10^{-9}$ & 1.45 && 2.00  $\times 10^{-8}$  & 0.62 && 2.29 $\times 10^{-8}$\\
4.11 && 1.02  $\times 10^{-8}$ & 1.86 && 2.14  $\times 10^{-8}$  & 0.97 && 2.57 $\times 10^{-8}$\\
5.70 && 1.32  $\times 10^{-8}$ & 3.63 && 2.66  $\times 10^{-8}$  & 1.53 && 3.60 $\times 10^{-8}$\\
8.66 && 3.08  $\times 10^{-8}$ & 5.94 && 5.37  $\times 10^{-8}$  & 2.99 && 4.32 $\times 10^{-8}$\\
\hline \hline
\end{tabular}
\caption{Fluid conductivities of hexadecane with AOT with different moisture content (listed as weight percentage).}
\label{cond}
\end{center}
\end{table*}

\section{Results and discussion}

\begin{figure}[h]
\centerline{\includegraphics[width=0.8 \linewidth]{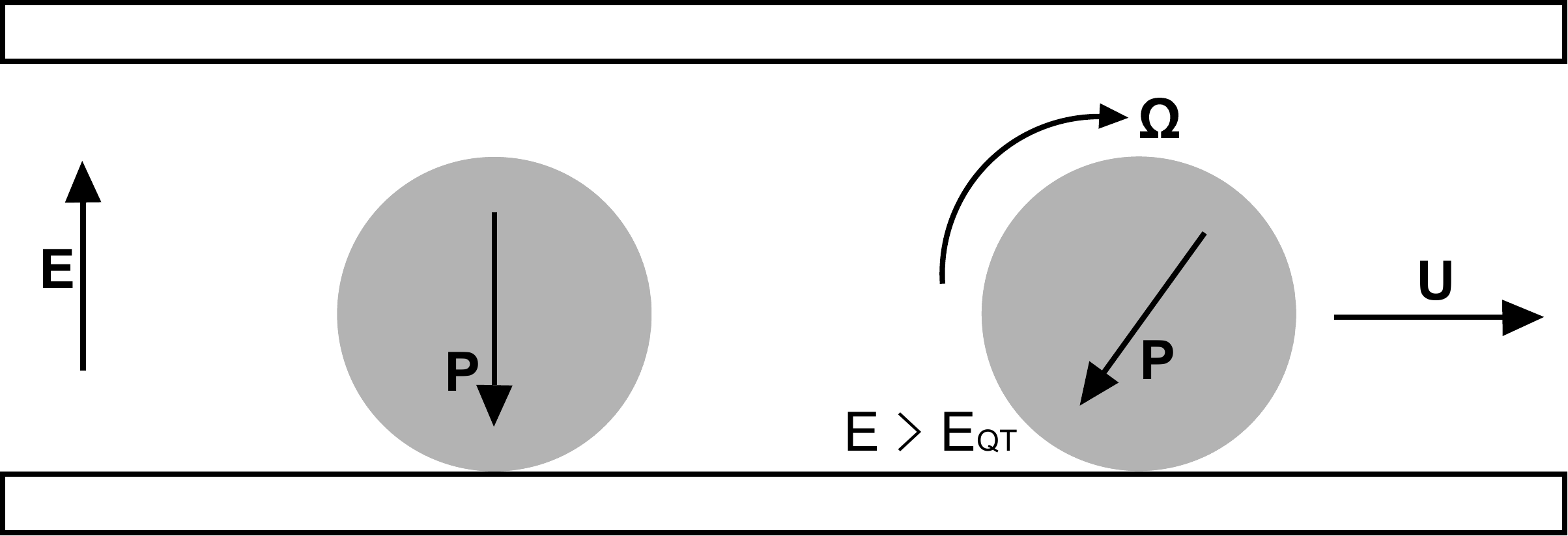}}
\centerline{\includegraphics[width=0.8\linewidth]{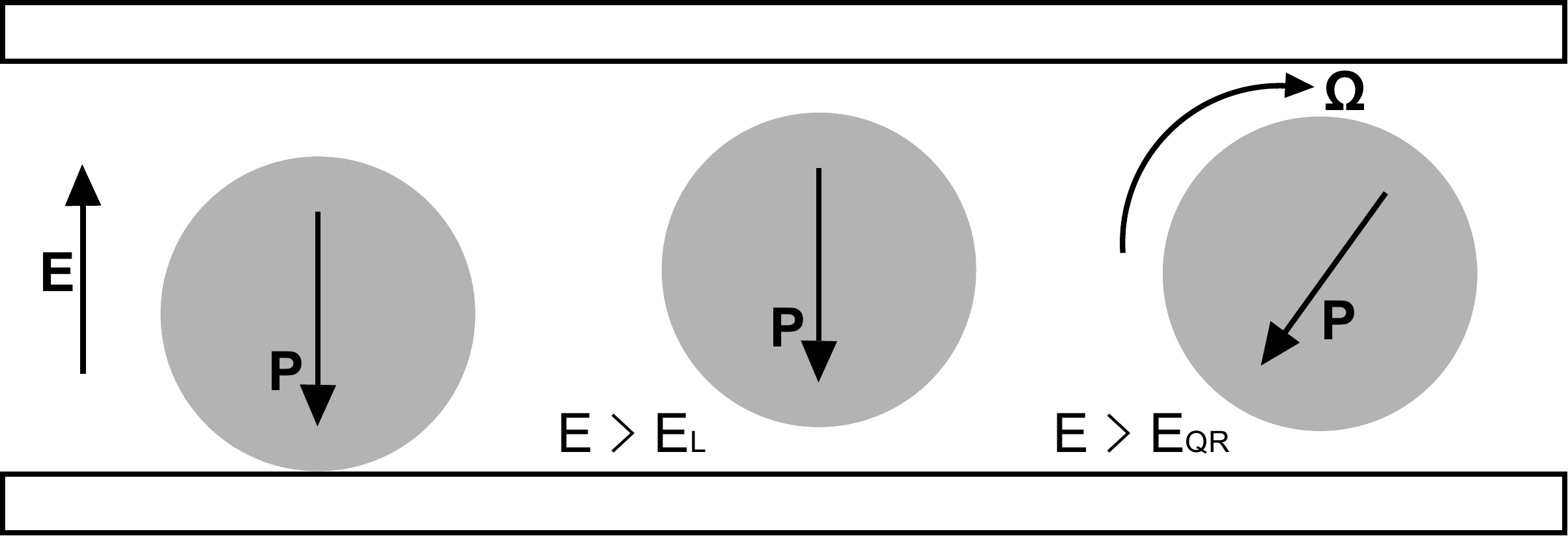}}
\caption{\footnotesize Illustration of the Quincke dynamics in the rolling (top) and hovering (bottom) regimes (see also Supplemental Videos.) 
}
\label{fig1a}
\end{figure}

We find two different particle dynamics
depending on the AOT and its water content {(see Fig.\ref{fig1a} for illustrations and  Supplemental Movies).} 
 In the moist AOT system, the sphere exhibits translational motion along the bottom chamber surface (called rolling \cite{Bartolo:2013}) above a threshold field strength $E_{QT}$ , as previously observed \cite{Bartolo:2013}. However, in the dry AOT system, the sphere first lifts off from the electrode  and levitates (without spinning) between  the electrodes. The lift occurs above a field strength $E_{L}$. Upon further increase in the field, rotation starts above a critical value $E_{QR}$.
 In general, $E_Q<E_L<E_{QR}<E_{QT}$.
  {$E_Q$ for the unconfined rotation is most sensitive to fluid viscosity and conductivity, as seen from \refeq{quinckeE}. Using a typical value for the hexadecane and AOT mixture, $\sigma_\out\sim 10^{-8}$ S/m,   yields $E_Q\sim 0.5$ MV/m.  Our measurements for the  electric field at onset of rolling $E_{QT}$ are in the range 1-5MV/m, consistent with the reported values by \cite{Bartolo:2013} and \cite{Lu:2018}. Lift and rotation in the moisture-free system $E_{QL}$ and $E_{QR}$ require lower fields compared to rolling,  0.5-1 MV/m.  Next we analyze the dependence of the critical fields for rolling ($E_{QT}$), lift ($E_{L}$),  and rotation in the levitated state ($E_{QR}$) on particle confinement, AOT concentration and water content. }
 
\subsection{Rolling}

\begin{figure}[h]
\centerline{\includegraphics[width=\linewidth]{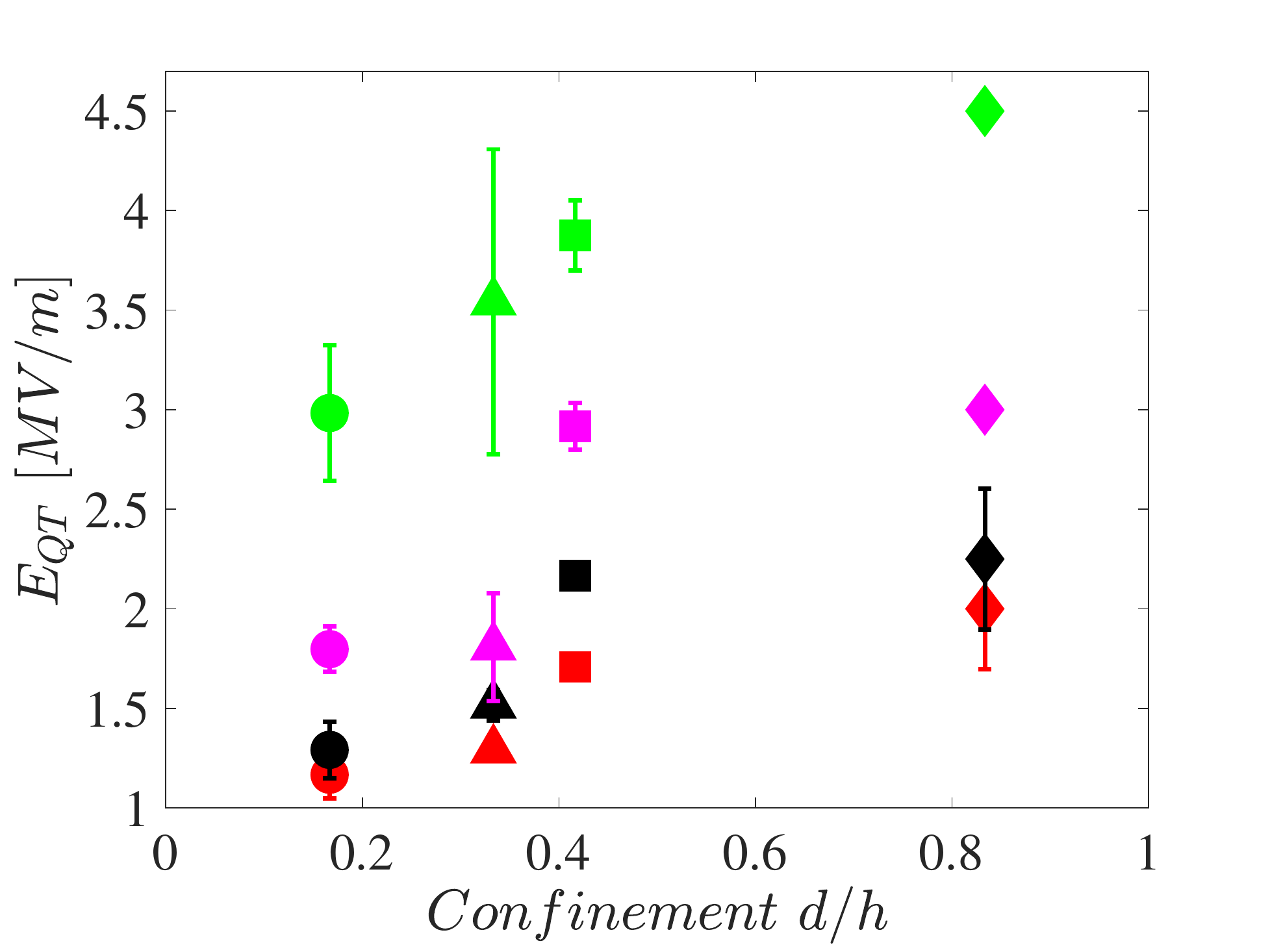}}
  \caption{\footnotesize Quincke rolling thresholds $ E_{QT}$ at different confinement for 0.1  M AOT in hexadecane and varying moisture content  $\%wt_{H_2O }$ =  red $<$ black $<$ magenta $<$ green (see Table \ref{cond}  for the values of the water weight percentages).  For comparison $E_Q$ for this system is 0.37MV/m in the absence of moisture. }
\label{fig3a}
\end{figure}

\begin{figure}[h]
\centerline{\includegraphics[width=\linewidth]{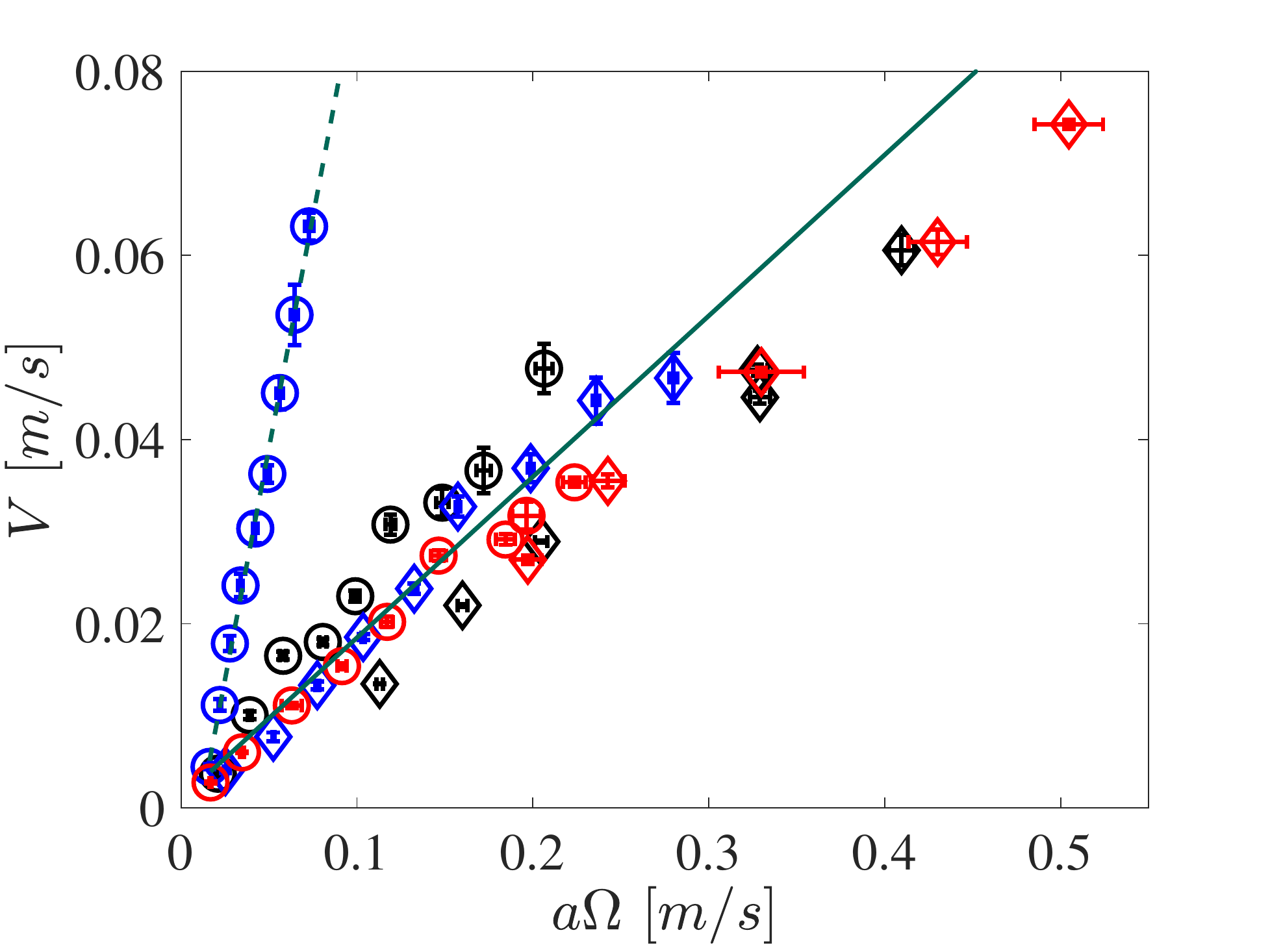}}
  \caption{\footnotesize Roller speed $V$ versus the no-slip rolling velocity   $ a\Omega$, $V=f(\delta) a \Omega$, for the lowest ($\circ$) and highest ($\diamond$) moisture contents  at 0.05 M (blue), 0.1 M (black), and 0.15(red) M AOT and $d/h = 0.83$. The dashed line corresponds to the case of rolling with no slip $f(\delta)=1$.  The solid line is the velocity corresponding to a $\delta$=10nm gap between the sphere and the bottom surface, calculated from \refeq{fd}.  }
\label{fig6}
\end{figure}

 Figure \ref{fig3a} shows that the electric  field strength above which the sphere starts to roll  increases with confinement.   
 One possible reason is the enhanced drag on the sphere due to the walls. To estimate the wall effect, we consider  a sphere with radius $a$
 translating with velocity $V$ and rotating with rate $\Omega$ near a wall. {The proximity of the wall modifies the force and torque on the sphere, compared to the unbounded case, 
 depending on  the separation between the sphere and bottom surfaces $\delta$:}
 $F^t=6 \pi \mu_\out a V f^t(\delta)$, $F^r=8 \pi \mu_\out a^2 \Omega f^r(\delta)$, $T^t=8 \pi \mu_\out a^2 V \tau^t(\delta)$ and $T^r=8 \pi \mu_\out a^3 \Omega \tau^r(\delta)$.
  For very small separations, the friction coefficients are derived from lubrication theory \cite{Goldman:1967a}
\begin{equation}
\label{eq1}
\begin{split}
f^r=&\textstyle{-\frac{2}{15}  \ln\left(\frac{\delta}{a}\right) -0.2526\,,\quad f^t=\frac{8}{15} \ln\left(\frac{\delta}{a}\right)-0.9588}\\
\tau^r&\textstyle{=\frac{2}{5} \ln\left(\frac{\delta}{a}\right)-0.3817\,,\quad \tau^t=-\frac{1}{10} \ln\left(\frac{\delta}{a}\right)-0.1895}
\end{split}
\end{equation}
The balance of forces and torques acting on the sphere is  $F^t+F^r=0$ and $T^E+T^r+T^t=0$, where $T^E$ is the electric torque exerted by the field.
  The   torque balance {shows that the electric field has to overcome a larger viscous  torque on the sphere compared to the unbounded case}
\begin{equation}
\begin{split}
T^E=8 \pi a^3 \mu \Omega \tau(\delta)\,,\\
 \tau \left(\delta\right)=-\tau^r(\delta)+\frac{f^r(\delta)}{f_t(\delta)}\tau^t(\delta)\,.
 \end{split}
\end{equation}
The gap between the sphere and the bottom surface $\delta$ is estimated to be few nanometers (see discussion of Figure \ref{fig6}). Using $\delta=10$ nm yields   $\tau\sim 4$; accordingly the critical field increases by a factor of 2, based on the linear relation between $E_Q^2$ and the viscous torque, \refeq{quinckeE}. This evaluation, however, ignores the top wall, whose effect can be non-negligible, especially at the highest confinement where the gap between the sphere and top electrode surfaces is below  20 $\mu$m. 

The separation between the translating sphere and the bottom surface can be estimated from the measured velocity slip, i.e., the difference between the translational velocity and the no-slip rolling velocity $a \Omega$. 
We experimentally measure the rotation rate  $\Omega$ by tracking surface features of the colloidal sphere (with radius $a=50\,\mu$m) using high speed video imaging. Figure \ref{fig6} compares the roller speed to the no-slip translational velocity $\Omega a$.  Only at the lowest AOT and moisture content does the sphere roll without slipping, i.e., $V= a \Omega$.
All the other cases show slip {$V< a \Omega$ depending} on the thickness of the lubricating film between the sphere and the electrode surface. 
Using the force balance on the sphere and the results from lubrication theory \refeq{eq1}, we find
\begin{equation}
\label{fd}
\begin{split}
 V=f(\delta) a \Omega\,,\quad 
f(\delta)=\frac{\frac{2}{15} \ln\left(\frac{\delta}{a}\right)+0.2526}{\frac{8}{15}  \ln\left(\frac{\delta}{a}\right)-0.9588}\,.
\end{split}
\end{equation}
From the data on Figure \ref{fig6}, we estimate the gap $\delta$ between the particle and electrode surfaces to be about ten nanometers.
 \begin{figure}[h]
\centerline{\includegraphics[width=\linewidth]{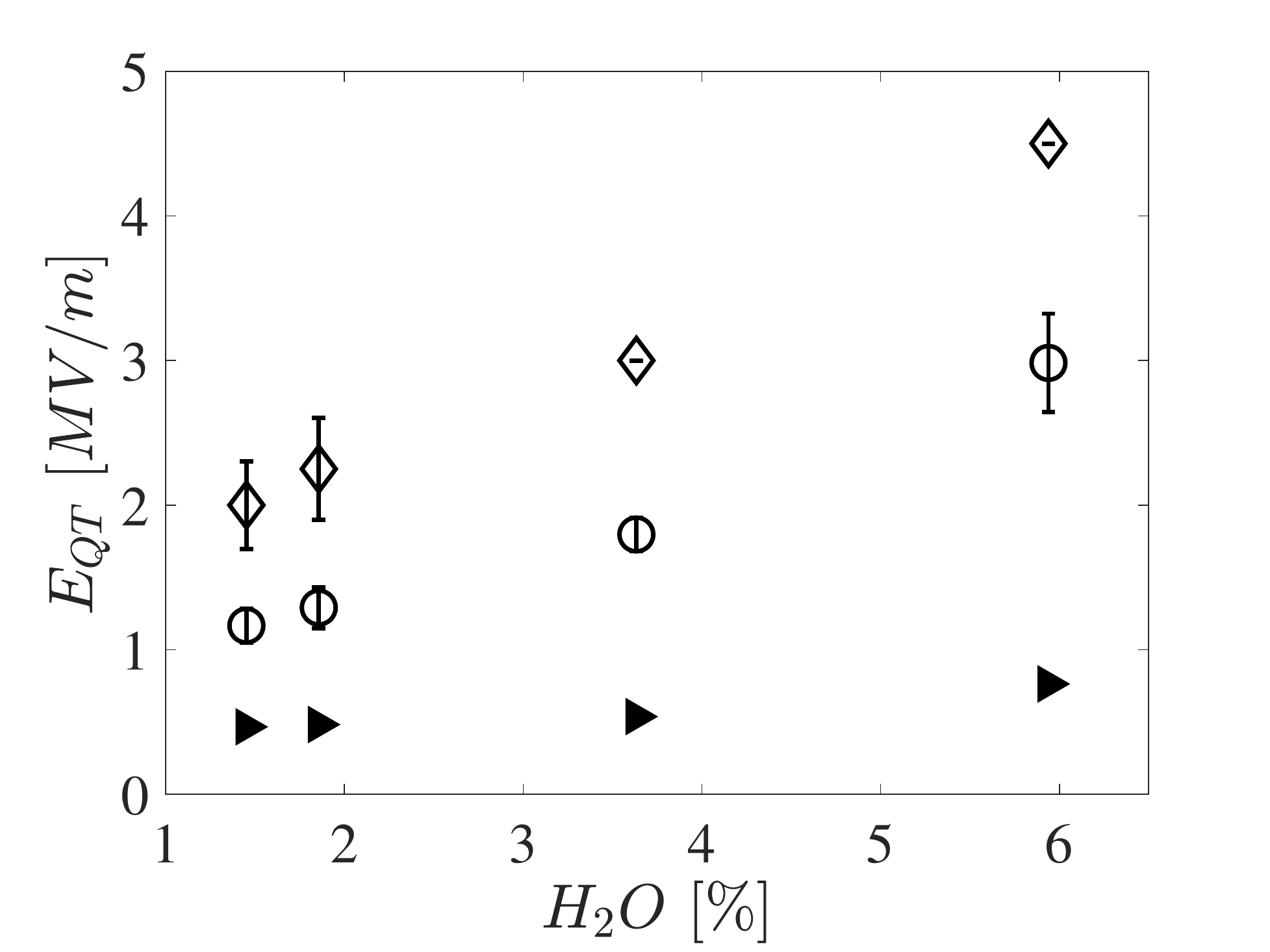}}
  \caption{\footnotesize  Quincke rolling thresholds for 0.1 M AOT-hexadecane solution as a function of moisture content. The symbols correspond to particle confinement $\diamond :  d/h = 0.83$, and $\bigcirc : d/h = 0.17$. 
   The solid symbols  are the threshold  for unconfined electrorotation calculated from \refeq{quinckeE} using the conductivities corresponding to the 0.1M  AOT and moisture content, Table \ref{cond}.}
\label{fig4}
\end{figure}
Another 
likely cause for the increase in the critical field for rolling is a resistance due to adhesion between the particle and the bottom surface. It has been reported that small amounts of water generate strong  adhesive force between surfaces in nonpolar fluids  \cite{Tombs:1995, Kanda:2001}. Furthermore, adsorption of water on the particle surface (e.g., in the form of AOT inverted micelles) increases the particle effective conductivity \cite{Tombs:1995,Eastoe:2015}. From \refeq{quinckeE}  we see that increasing the conductivity ratio $\Rr$ decreases the denominator and hence increases $E_Q$.
 Indeed, we observe that 
 at a given AOT concentration, higher moisture increases the critical field for rolling, see Figure \ref{fig4}.

\begin{figure}[h]
\centerline{\includegraphics[width=\linewidth]{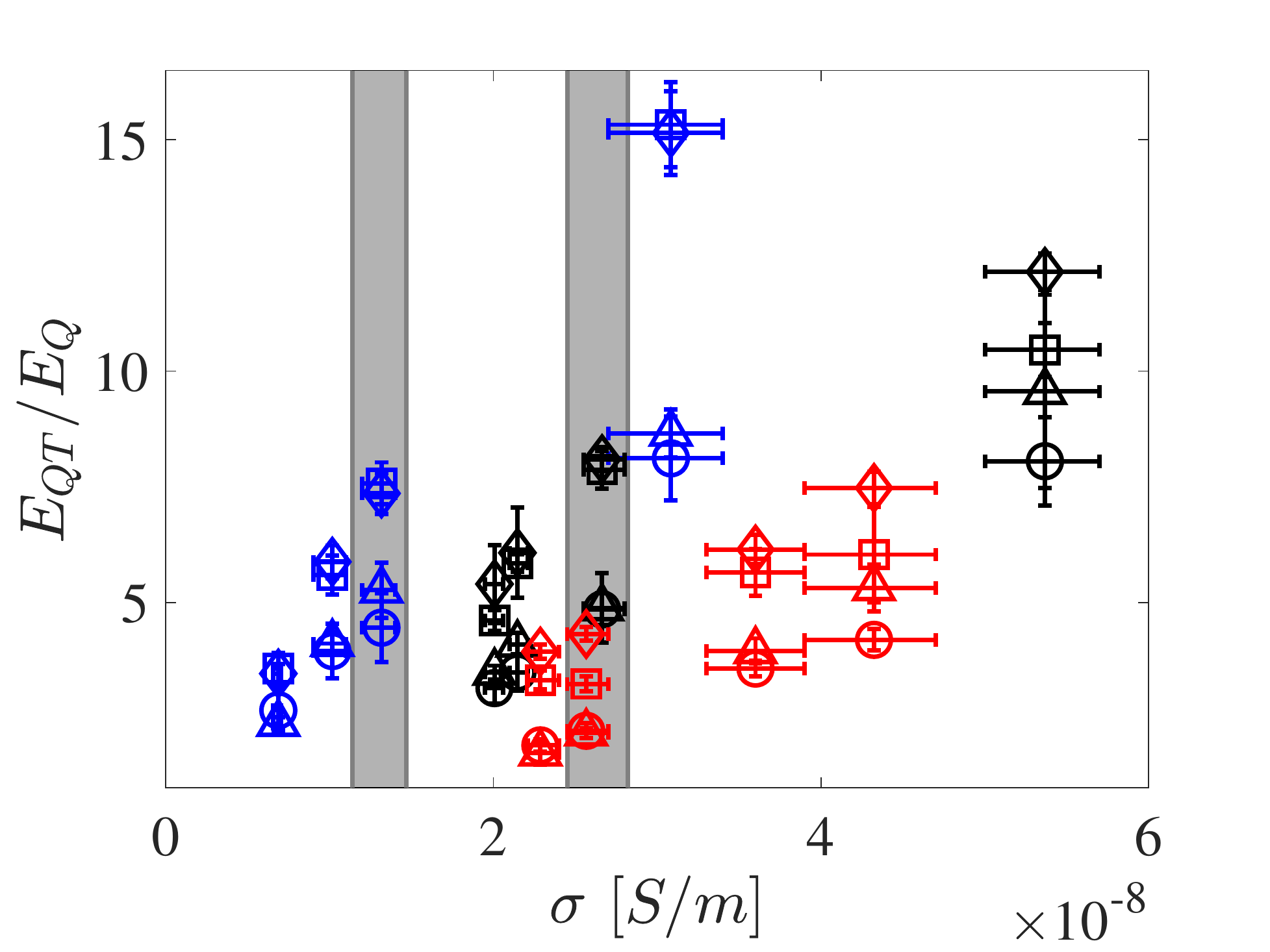}}
  \caption{\footnotesize Quincke rolling threshold as a function of conductivity of the suspending fluid. The symbols correspond to different confinement $\diamond :  d/h = 0.83$, $\square : d/h = 0.42$, $\triangle : d /h = 0.33$, $\bigcirc : d/h = 0.17$.  Colors denote AOT concentration 0.05 M (blue), 0.1 M (black) and 0.15 M (red). The electric field is scaled by the Quincke threshold for unconfined rotation calculated from \refeq{quinckeE} for the conductivity of the moisture-free fluid. The regions shaded in grey show that even for  solutions with similar conductivities and same confinement,  the $E_{QT}$ can vary due to different moisture content.}
\label{fig3}
\end{figure}

Figure \ref{fig3}  shows that, in general,  the threshold  for rolling increases with fluid conductivity as expected from the behavior at unconfined electrorotation \refeq{quinckeE}.  However,  unlike the unconfined rotation, there is no  unique relation between the threshold field and  the conductivity of the suspending fluid, as highlighted by  the shaded regions on Figure \ref{fig3}.
This is related to 
fact that same conductivity may correspond to different 
moisture content, see Table \ref{cond}.

\begin{figure}[h]
\centerline{\includegraphics[width=\linewidth]{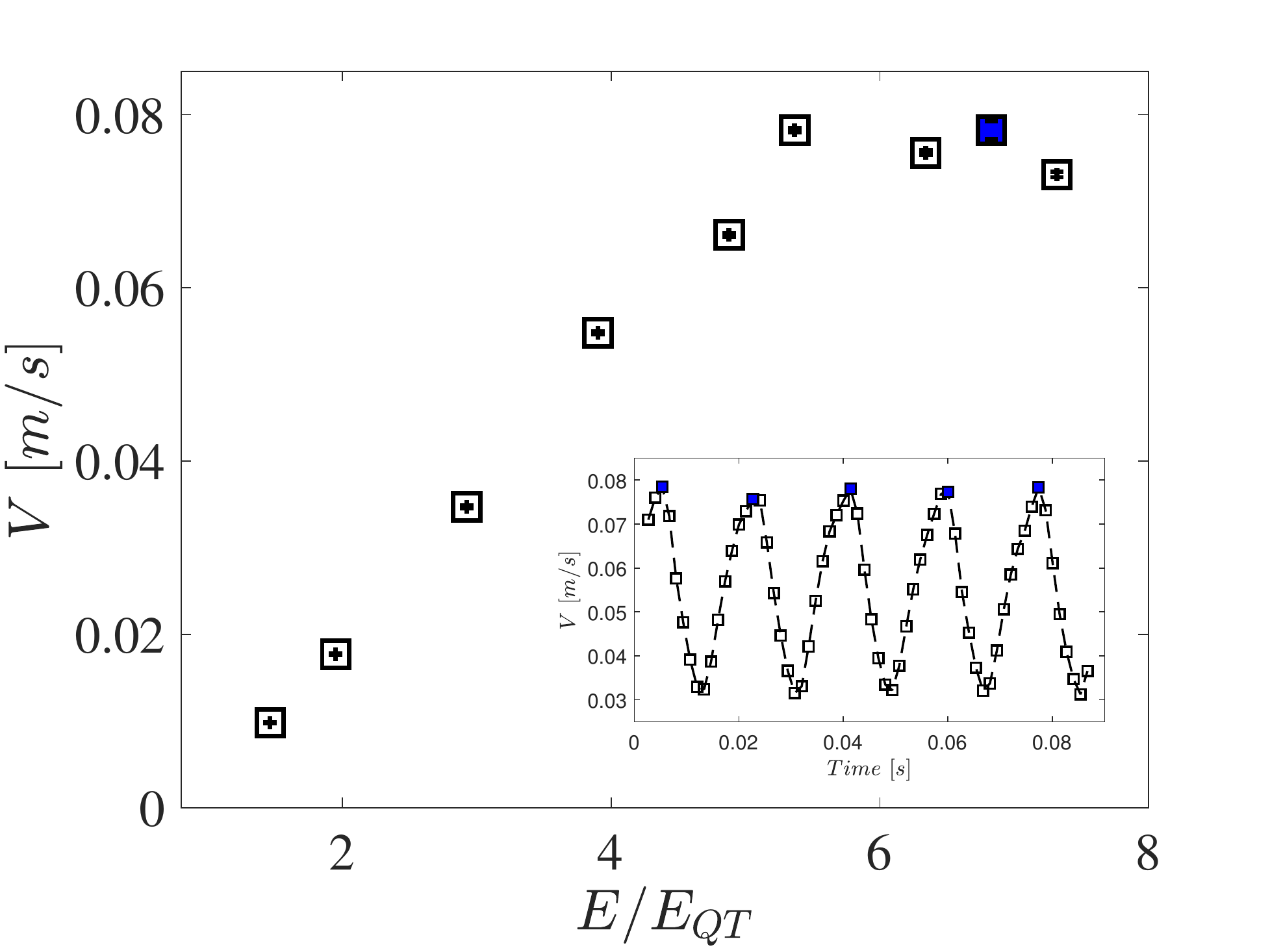}}
  \caption{\footnotesize Quincke roller velocity as a function of field strength $E/E_{QT}$, where $E_{QT}$ is the critical field for the onset of rolling for the system  0.1 M AOT with 1.45$\%$  water
 and confinement $\square\, d/h = 0.42$.  
  }
\label{fig5}
\end{figure}

The classical  Quincke effect results in constant torque and suggests rolling with a constant velocity which increases linearly with field strength for $E/E_{QT}\gg 1$, see \refeq{quinckeW}. 
 Figure \ref{fig5} shows that the rolling velocity  follows the expected  dependence on the field strength. 
At high field strengths, however, we find a previously unobserved dynamics - the particle undergoes periodic acceleration and deceleration, see the inset in Figure \ref{fig5}, and the average velocity plateaus. The new unsteady dynamics likely arises from particle inertia: the rolling is very fast, $V\sim 0.1$ m/s, and the particle Reynolds number becomes $Re=\rho_\out V a/\mu_\out \sim 1$. The  steady rotation, \refeq{quinckeW}, is predicted
under the assumption of negligible particle inertia \cite{Lemaire:2002}.   The Quincke model with inertia maps onto the Lorenz  equations, which can have periodic and chaotic solutions, corresponding to unsteady rotations \cite{Lemaire:2005}.  
\vspace{-10pt}
\subsection{Hovering}

In the case where the AOT added to the hexadecane is anhydrous,
 the sphere first lifts off the electrode at a critical field strength $E_L$ and starts to rotate, while hovering in the space between the electrodes, above a  threshold field $E_{QR}$. 
 \begin{figure}[h]
\centerline{\includegraphics[width=\linewidth]{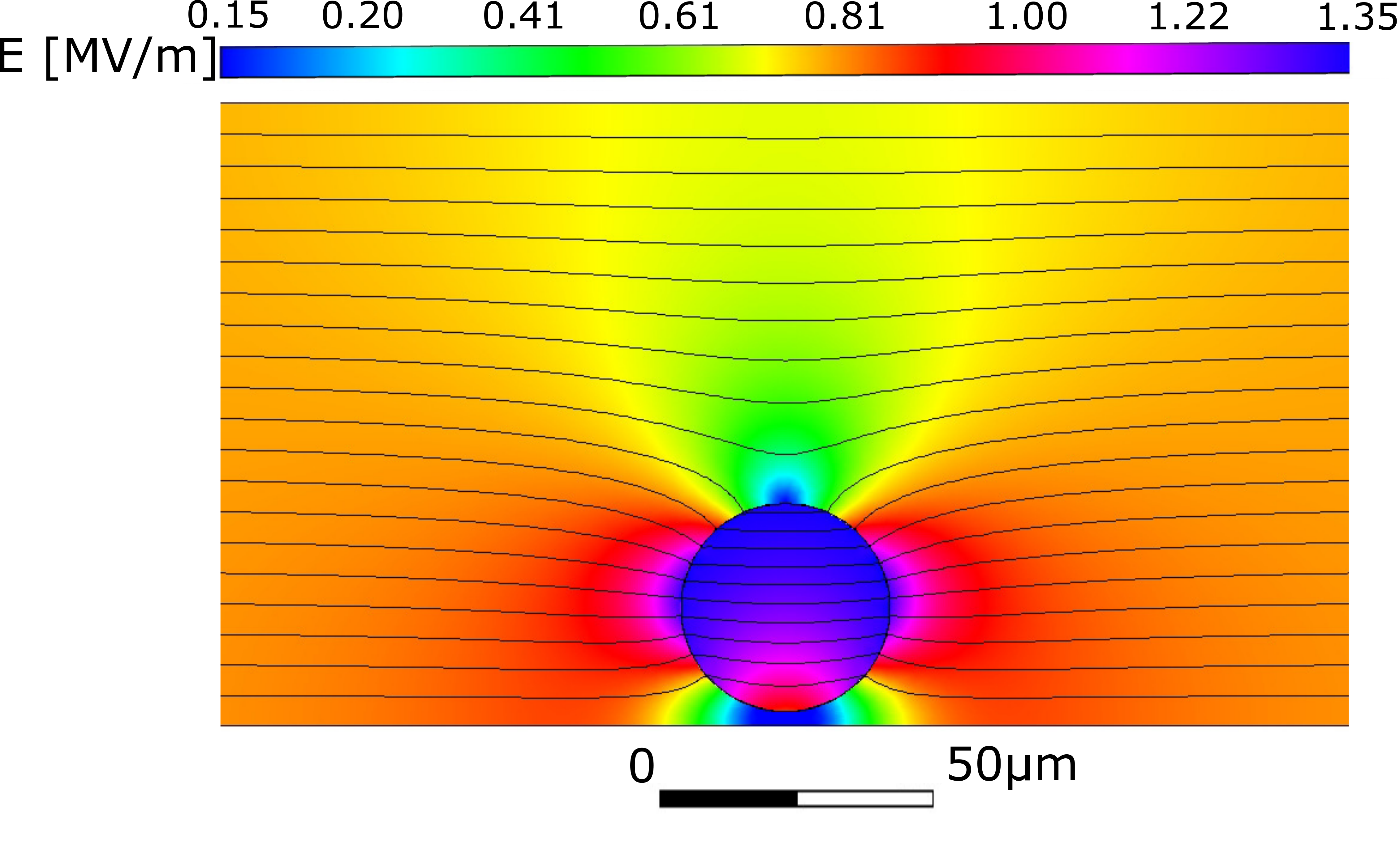}}
\caption{ \footnotesize Electric field  magnitude (color) and potential lines about a sphere with $d=40\, \mu$m
between two parallel electrodes separated by a 120 $\mu m$ gap. Particle  conductivity is $\sigma_\ins = 10^{-17}$ S/m 
and the  medium conductivity is $\sigma_\out = 2\times 10^{-8}$S/m. The potential difference is 90V and the particle is suspended at a height of 2.8 $\mu m$ above the bottom electrode.  Computations performed with ANSYS-FLUENT.}
\label{fig7}
\end{figure} 
  The lift is of dielectrophoretic origin, due to the highly nonuniform electric field around the confined sphere  \cite{Young:2005, Hossan:2013}, illustrated in Figure \ref{fig7}. {The repulsion is somewhat counterintuitive because approximating the sphere as a point dipole predicts attraction to the wall (the effect of the equipotential electrode surface in a uniform electric field is equivalent to an image dipole also antiparallel to the field  \cite{Bartolo:2013}). The image argument ignores the nonuniformity in the electric field induced by the particle, which is reasonable for small particles compared to the electrode separation as in the experiments  \cite{Bartolo:2013}. However, in our system  the perturbation in the applied electric field by the sphere is large and top-bottom asymmetric. The  surface charge distribution is also asymmetric
Thus,  effectively each half of the dipole ``feels" a field of different magnitude  leading to a  net force on the sphere.}
 The direction of this
  force depends on the particle and suspending medium conductivities and can be estimated from the particle dipole. In the Quincke configuration, for  the PMMA sphere,
{the dipole points in the opposite direction to the electric field} $\sim (\Rr-1)<0$ and the particle moves  away from high field regions {near the bottom electrode}, towards the middle of the chamber.

The lift is suppressed in the moist AOT system likely due to either adhesion between the particle  and the bottom surface and/or modified particle conductivity due to adsorption of water (e.g., in the form of AOT inverted micelles) \cite{Tombs:1993, Tombs:1995, Kanda:2001, Eastoe:2015}. In the latter case, the formation of highly conducting layer even if of  nanometric thickness  increases the effective conductivity of the particle and may reverse the sign of the particle induced dipole thereby changing the dielectrophoretic force from repulsive to attractive.

\begin{figure}[h]
\centerline{\includegraphics[width=\linewidth]{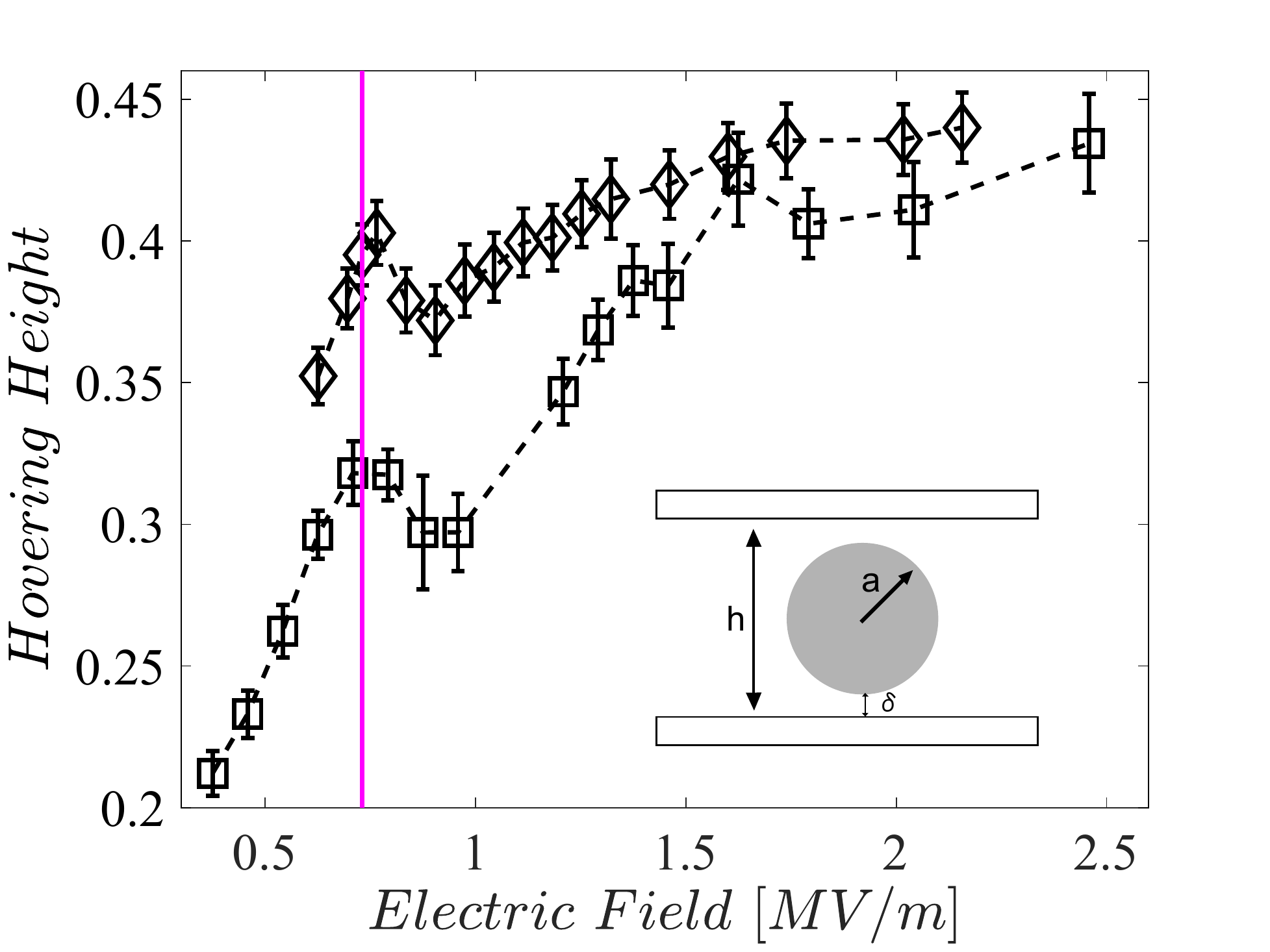}}
  \caption{\footnotesize 
Hovering height scaled by the distance between the electrodes, $(a+\delta)/h$,
  as a function of electric field for a 0.15M dry AOT and confinement $\diamond\,  d/h = 0.83$ and $\square\, d/h = 0.42$. The onset of rotation  is marked by the magenta line. {Hovering height of 0.5 corresponds to the midplane between the electrodes.}
}
\label{fig8}
\end{figure}

After lift-off, the hovering height increases with field strength. {We experimentally measure the height directly  using the microscope focus knob with custom designed calibration which enables 1.24 $\mu$m  resolution}.Figure \ref{fig8} shows that the equilibrium height, defined as the distance between the sphere center and the bottom surface,  approaches the chamber midplane as the field increases. At the onset of rotation, the equilibrium height decreases slightly because the dipole tilt decreases the dipole component antiparallel to the applied field direction 
thereby effectively 
decreasing the dielectrophoretic lift force. The hovering height is sensitive to confinement. The most confined particle experiences the strongest initial lift, likely due to electric field gradients being largest in this case. {The hydrodynamic interaction of the rotating sphere and the confining electrode surfaces induces particle translation, however the effect is much weaker compared to the rolling case \cite{Swan:2007,Goldman:1967a, Driscoll:2017}: the translation is two orders of magnitude slower than rolling.}

\begin{figure}[h]
\centerline{\includegraphics[width=\linewidth]{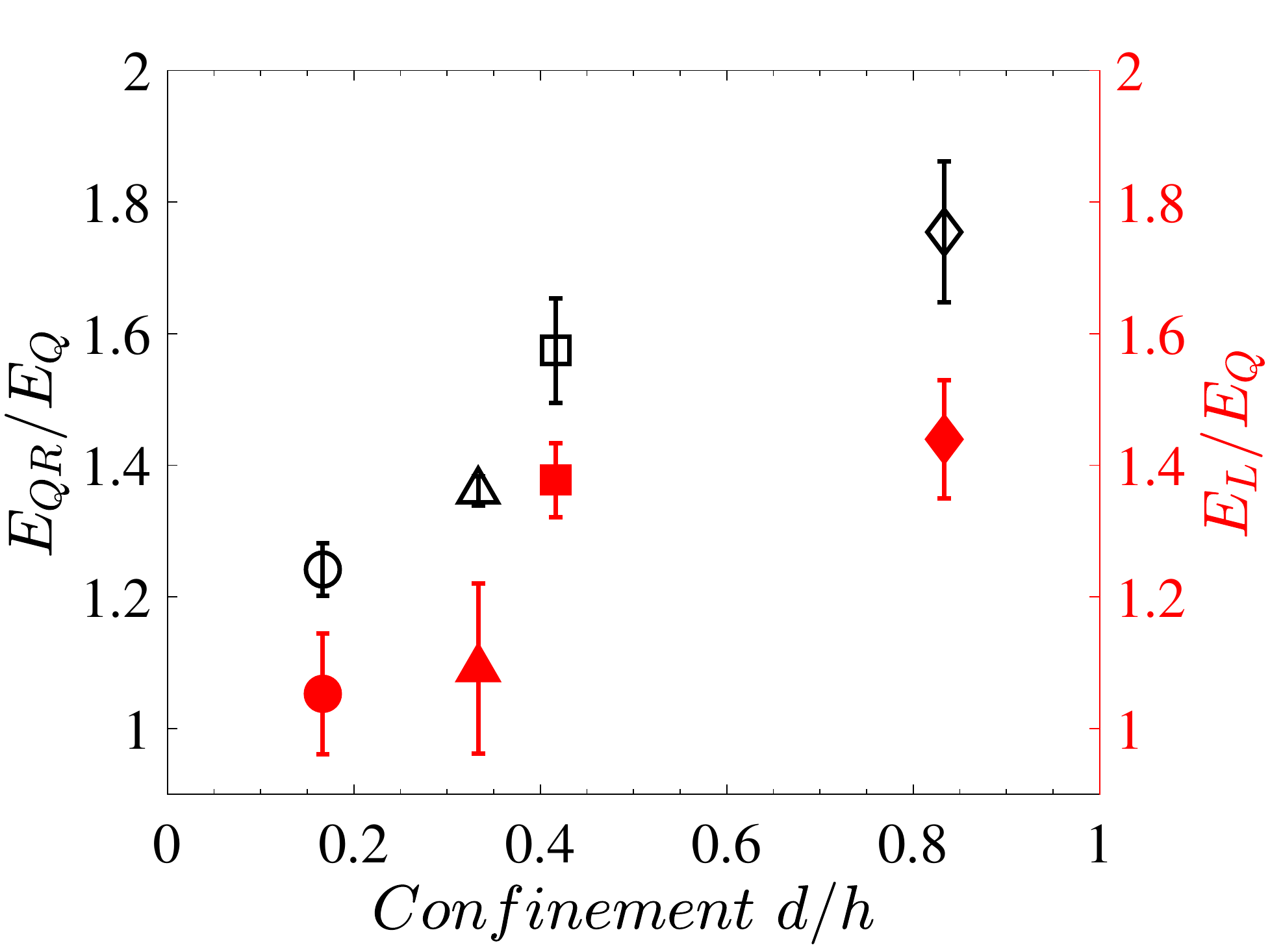}}
  \caption{\footnotesize Thresholds for lift (right vertical axis) and rotation (left vertical axis) as a function of confinement for a 0.1 M dry AOT. The thresholds have been non-dimensionalized by the threshold for Quincke rotation in unbounded medium calculated from \refeq{quinckeE}  using the conductivity of the dry fluid of the same AOT concentration (0.1 M). 
}
\label{fig9}
\end{figure}

Figure \ref{fig9} shows that the threshold for rotation   $E_{QR}$ increases, albeit modestly,  with confinement. However, the threshold for rotation is lower than the rolling case (at the highest confinement the rotation onset is higher than the unconfined case by a factor of 2 while in the rolling case with highest moisture the factor was about 10).  This suggests that even though  confinement results in stronger viscous resistance to the sphere rotation, water effects (either resulting in adhesion or increased effective conductivity) play a more important role in the rolling onset. 

\begin{figure}[h]
\centerline{\includegraphics[width=\linewidth]{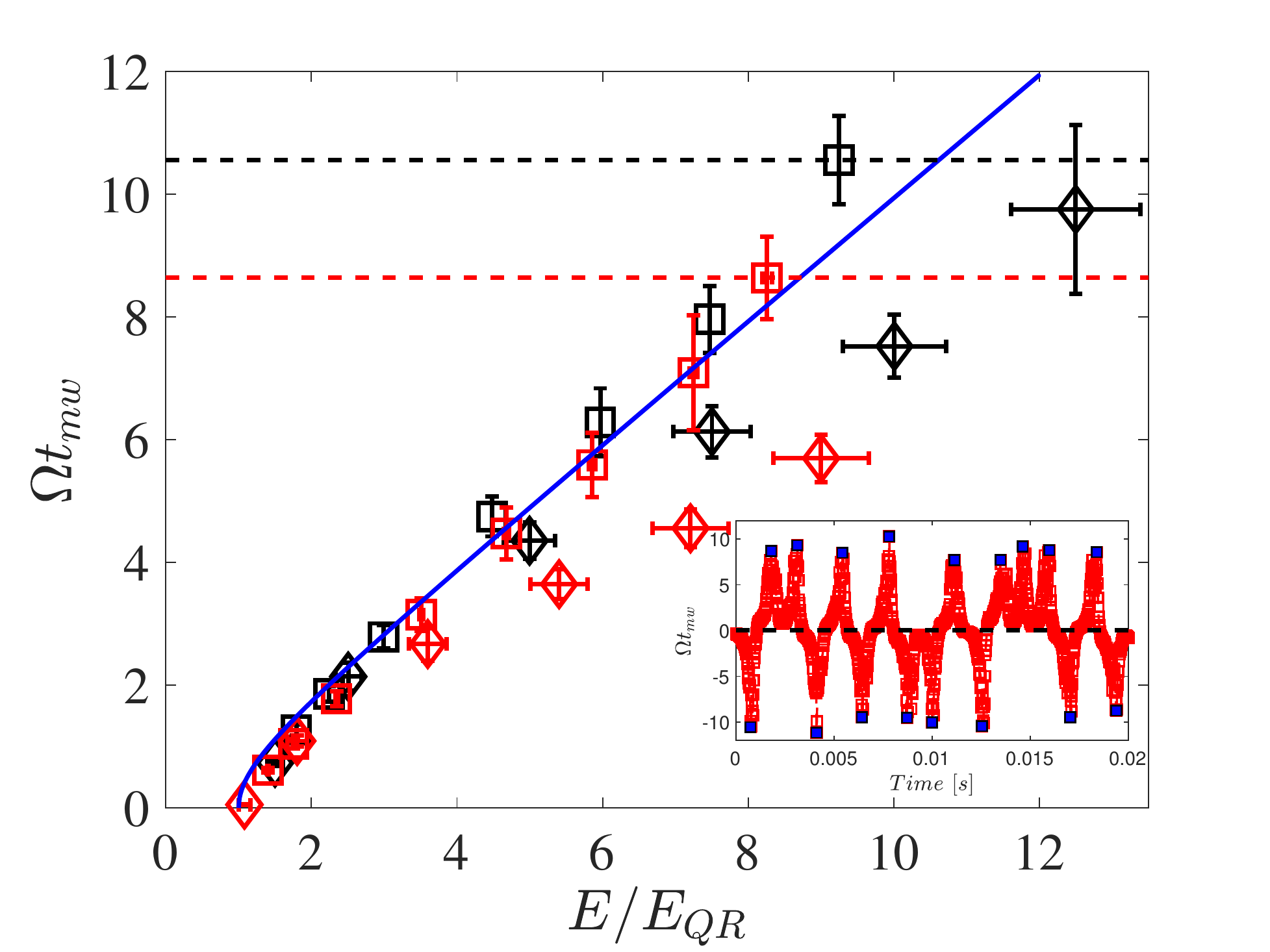}}
  \caption{\footnotesize Rotation rate of the hovering Quincke rotor for 0.1 M (black) and 0.15 M(red) AOT. The confinements  are $\diamond :  d/h = 0.83$, and $\square : d/h = 0.42$, The solid line is obtained from \refeq{quinckeW} using the experimentally determined $E_{QR}$ instead of $E_Q$. The onset of the chaotic regime is marked by the  black and red horizontal lines for 0.1 M and 0.15M respectively. The inset illustrates the chaos in rotation.}
\label{fig10}
\end{figure}

Figure \ref{fig10} shows the experimentally measured rotation rate in the hovering state as a function of the electric field strength.  Since  the rotation rate  $\Omega$ is measured by tracking surface features, only the larger sphere with diameter 100 $\mu$m is studied because   the $d=$ 40 $\mu$m particles  are featureless. Our experimental measurements for confinement $\frac{d}{h} = 0.42$, are in close agreement to those theoretically predicted using \refeq{quinckeW} using the experimentally determined $E_{QR}$ instead of $E_Q$. Increasing confinement  results in a decreased $\Omega$ due to additional hydrodynamic drag from the wall. 

In strong fields the rotor dynamics becomes chaotic due to particle inertia. During steady rotation the rotation axis is fixed, but in the chaotic case the rotation axis changes irregularly (while still remaining  in the plane perpendicular to the field direction).  The chaos in Quincke rotation has been previously 
studied in a 2D case (cylinder) \cite{Lemaire:2002, Lemaire:2005}. In this case the rotation axis is fixed and the chaos was manifested by the rotation rate randomly switching between clockwise and counterclockwise direction.  

\section{Conclusion}
We experimentally study the Quincke effect
(spontaneous spinning of a sphere  in a uniform electric field)  in strong confinement  with particle diameter to gap ratio $d/h$ ranging between 0.17 and 0.83. Our results quantify that confinement increases the critical field for onset of rotation, and thus, unlike the classical unbounded Quincke rotation, the threshold filed becomes dependent on the particle size.  

We find that the Quincke effect in confinement is very sensitive to the additive used to control the conductivity of the suspending oil. In our system of hexadecane with added AOT, moisture in the AOT dramatically changes the Quincke behavior. 

If the AOT contains water, a sphere initially resting at the bottom electrode adheres to the surface and above the Quincke threshold the sphere starts rolling. The adhesion appears stronger at higher water content, suggested by the experimentally observed higher field threshold for rolling. Comparing the roller translational velocity and the no-slip rolling velocity calculated to the rotation rate shows that the Quincke rollers roll with slip.
Mixtures of water and AOT can  have similar conductivity but different threshold for rolling, confirming the important role played by water in the adhesion.

If the added AOT is anhydrous, adhesion is prevented, the sphere lifts off from the electrode surface due to electrostatic repulsion  and rotates while hovering in the space between the electrodes. The critical fields for lift and rotation in this case are an order of magnitude lower than the rolling case. 

In stronger fields, rolling becomes unsteady with time-periodic velocity, while rotation in the hovering state becomes chaotic. 

Our study highlights the complex dynamics of the Quincke effect. Given the increasing interest in the Quincke rotors as a model ``active" particle (either self-propelled in the rolling case or self-rotating in the hovering case), our study provides important insights about how to harness the Quincke  effect for active fluids. 

\section{Acknowledgements}
This research was funded in part by NSF awards CBET-1704996 and CMMI- 1740011.

\bibliographystyle{unsrt}

\end{document}